\documentclass[journal]{IEEEtran}
\usepackage[english]{babel}
\usepackage{nicematrix}
\usepackage[top=2cm,bottom=2cm,left=1.5cm,right=1.5cm,marginparwidth=1.75cm]{geometry}

\usepackage{graphicx,subfigure,mhchem,amsmath,cite}
\usepackage[colorlinks=true, allcolors=blue]{hyperref}

\title{Dynamics of capillary effects \\in spin conversion of water isomers}
\author{Serge~Kernbach$^1$, Sergey M.~Pershin$^2$ \\[3mm]
\small $^1$CYBRES GmbH, Research Center of Advanced Robotics and Environmental Science,\\
\small Melunerstr. 40, 70569 Stuttgart, Germany, {\it serge.kernbach@cybertronica.de.com}\\
\small $^2$Prokhorov General Physics Institute, Russian Academy of Sciences, ul. Vavilova 38, Moscow, 119991, Russia
\vspace{-6mm}
}

\begin{document}
\maketitle
\thispagestyle{empty}

\begin{abstract}
This work explores the dynamics of capillary effects in pure \ce{H_2O} excited by hydrodynamic cavitation, which has been reported to introduce a non-equilibrium state of para- and ortho- isomers of water.	Differential measurements are conducted with 0.3mm and 0.5mm capillary tubes and precision digital vernier height gauges. Water samples are degassed at -0.09 MPa, their temperature is equalized. Since the non-equilibrium state of isomers is expected to exist in ice-like building blocks near the water interfaces, the immersion depth of capillary tubes was set to 10mm, 2mm and about 0.1mm below the surface. The measurement results show the difference in surface tension of 6.7\%-11.3\% with a maximum value of about 15.7\% between control and experimental samples. These effects in near-surface layers are observed within 30-60 minutes after the excitation. Experimental samples taken from the surface demonstrate movement about 2 times longer than control ones; their values are largest in the whole series of experiments. Micro-manipulation tool recorded an interesting effect of stopping the liquid in capillary tubes, observed mostly in experimental samples, where the meniscus and surface tension oscillated at 0.05-0.1mm level. The described approach has two main applications. First, the measurement system can be used as the fast and cost effective detector of spin-based phenomena that affect viscosity and surface tension. Second, since the capillary effects plays an important role in aquaporin channels and plant water transportation system, these techniques can be applied in phytosensing, in particular, in sap flow measurements, and in a hydroponic production.
\end{abstract}


\section{Introduction}

Capillary effects play an important role in biological systems such as aquaporin channels \cite{Murata00}, plant water transportation \cite{10.1104/pp.15.00333} or blood circulation \cite{ARTMANN19983179}, as well as in a number of application areas from irrigation and horticulture \cite{horticulturae4030023} up to lubrication \cite{STREATOR2002121}. Several publications \cite{Pershin09, kernbach2022electrochemical} demonstrated that the spin conversion of water isomers changes not only the electrochemical reactivity \cite{Kilaj18} but also the surface tension. Thus, it is expected that spin-based mechanisms can have essential biological and engineering implications in cases of capillary movements.

This work explores the capillary dynamics of pure \ce{H_2O} excited by hydrodynamic cavitation. Optical and NMR spectroscopy confirmed that the spin conversion process from equilibrium to non-equilibrium state takes place in such systems \cite{Pershin09NMR}. In particular, it was estimated that the cavitation enriches the water content with ortho- isomers by 12\%-15\% \cite{Pershin22} due to the collapse of cavitation bubbles when water passes through the supercritical state. For experiments we use 0.3mm and 0.5mm laboratory capillary tubes with differential measurement by precision vernier height gauges. Water samples undergo standard in such cases degassing and temperature equalization steps before measurements. 

Non-equilibrium state of water isomers is expected to exist in ice-like structures near the water interfaces \cite{Davis12,Pershin12ICE} and especially at the air-water interface \cite{Odendahl22}, which are intensively discussed in the community \cite{PershinIce10,PhysRevLett.101.036101,Wei01,en8099383,Monserrat20}. Experiments with evaporation of water \cite{Novikov10,POULOSE2023814} contribute to the discussion about non-equilibrium dynamics of spin isomers. To investigate this issue, we used the micro-manipulation tool to sample the surface area. Due to variation in production, sharp and round edges of capillary tubes have different geometries and thus different surface behaviour; in the worst case, samples are taken from an immersion depth of about 0.1mm. Results are compared with 2mm and 10mm immersion depth.

Experiments within this paper have been conducted in parallel with exploration of thermal effects caused by a heat capacity of water isomers \cite{kernbach23Thermal}, and follow the already published Ref \cite{kernbach2022electrochemical} that explored the electrochemical reactivity of isomers. These works provide experimental arguments towards spin conversion, predicted and observed in the gas phase \cite{Miani04,POULOSE2023814,Kilaj18,Tikhonov02}, which also takes place in liquid phase as reported earlier \cite{Pershin14}.

This approach has two main applications. The height $h$ of a liquid column is given by
\begin{equation}
\label{eq:liquidColumn}
h=\frac{2\gamma\cos{\theta}}{pgr},	
\end{equation}
where $\gamma$ is the liquid-air surface tension, $\theta$ is the contact angle, $p$ is the density of liquid, $g$ is the local acceleration due to gravity, and $r$ is the radius of tube. Temperature has a non-trivial effect on this dependency (see, e.g, the discussion in \cite{Grant02,She98}), thus both samples in differential measurements should have an equal temperature. However, the temperature dependency in the range 20-25$^\circ C$ is small (for instance, $\cos{\theta}$ changes in this range between 1 and 0.98 \cite{She98}). Since no other dependencies are involved in (\ref{eq:liquidColumn}), the capillary sensor can represent a simple and cost-effective detector for spin phenomena that affect viscosity and surface tension; at the same time, the sensor is resistant to variations of environmental parameters. Since the water transportation system in plants involves multiple capillary effects in stem, roots and leaves \cite{Martinez14, 10.3389/fpls.2021.615457}, the second application targets agriculture and phytosensing, in particular the sap flow measurements \cite{WatchPlant21} and hydroponic production.

\section{Methods and Setup}

\begin{figure}
\centering
\subfigure[]{\includegraphics[width=0.49\textwidth]{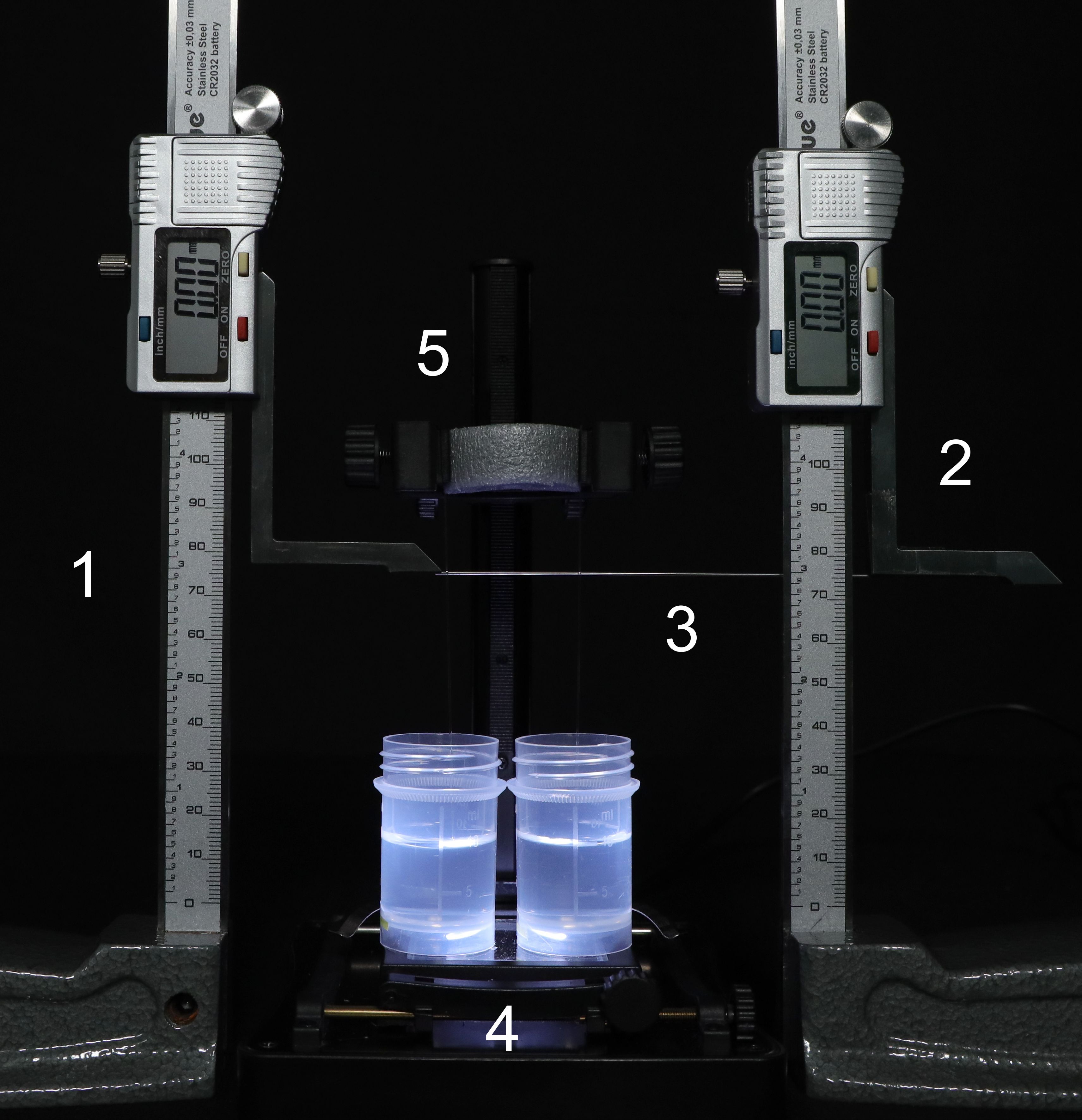}}
\subfigure[]{\includegraphics[width=0.49\textwidth]{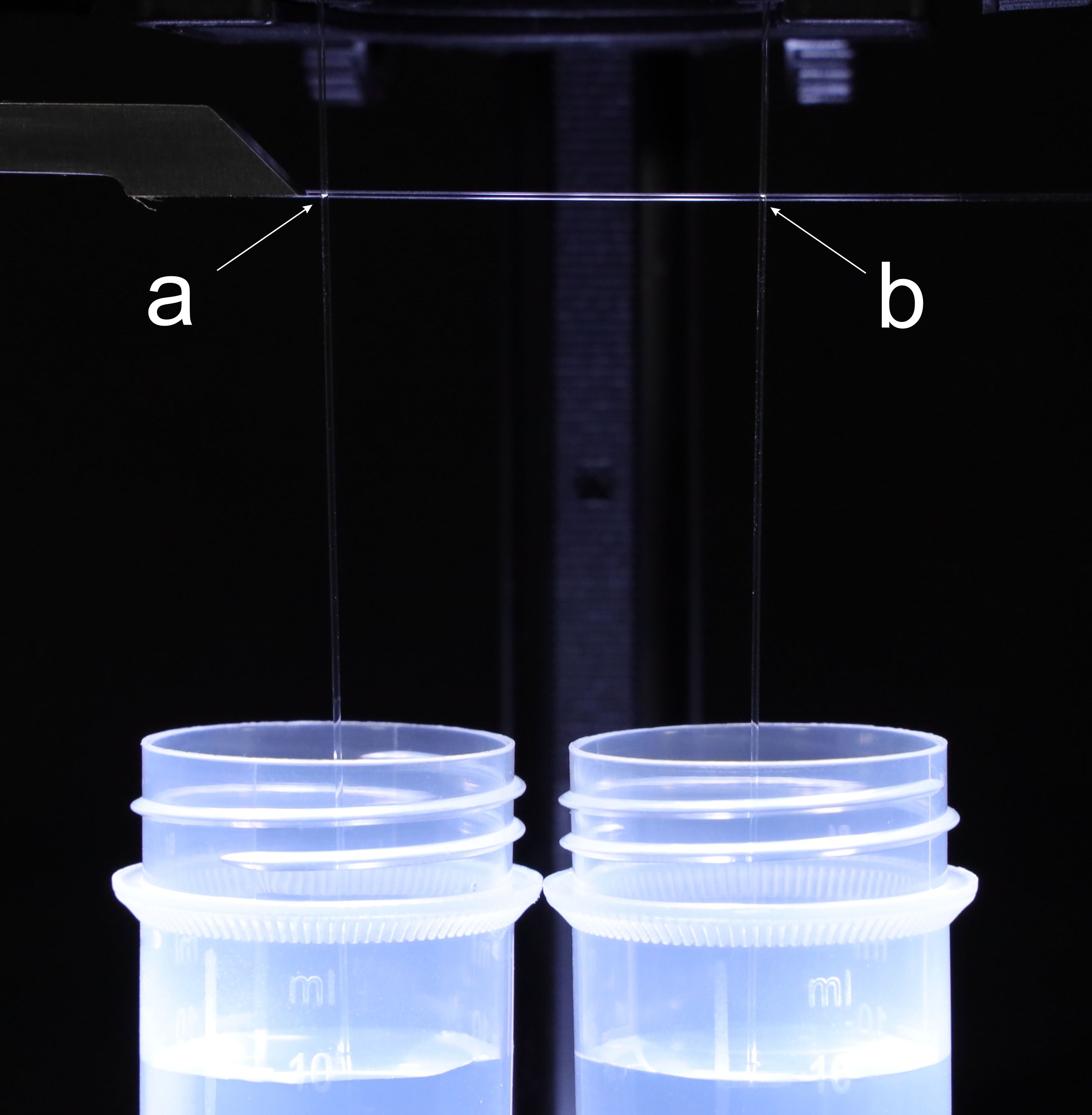}}
\caption{ \textbf{(a)} Measurement setup: 1, 2 -- digital vernier height gauges for experimental and control samples with accuracy $\pm 0.03mm$, 3 -- horizontal line for calibrating zero level on the control sample, 4 -- positioning table for samples, 5 -- lift for vertical movement of capillary tubes; \textbf{(b)} positions of meniscus for a -- experimental and b -- control samples. Example is shown with 0.3mm capillary tubes. \label{fig:setup}}
\end{figure}

\textbf{Measurements.} Capillary effects are observed in 0.3mm and 0.5mm glass tubes (pempered glass, 100 mm hight), placed vertically in open-surface containers with water. Differential measurements with two tubes are used, one sample represents an experimental channel, another one is a control. Immersing into the fluids is performed synchronously, both tubes are equalized in the vertical placement. Position of the meniscus is measured by two digital vernier height gauges with the measurement accuracy of $\pm0.03$mm, see Fig.\ref{fig:setup}. To estimate the dynamics of capillary effects, three readings are recorded: the first one after immersing into fluids, the second and third ones -- 3 and 6 min after. Each attempt has several independent measurements, from them the means and StDev are calculated and collected into the table of results.

\textbf{Preparation of samples.} Double distilled water with initial conductivity of $<0.05\mu S/cm$ is used. Experimental sample after treatment, together with the control sample is first degassed at -0.09 MPa in wide glass containers (large surface area with a liquid height of about 2mm). No mechanical or magnetic stirring is applied on this step to avoid excitation of the control sample. After this, the liquids are filled into hermetically closed containers with laboratory pipette (accuracy of filling 0.05 ml) and then are thermally equalized in a water bath for 7-10 min. Temperature of both fluids is controlled by a thermocouple sensor with accuracy about $\pm0.1^\circ C$. Two series of control and experimental samples are prepared at each attempt: 10 ml samples for measuring capillary effects and 15 ml samples for measuring thermal effects (conducted within \cite{kernbach23Thermal}).

\begin{figure}
\centering
\subfigure{\includegraphics[width=0.45\textwidth]{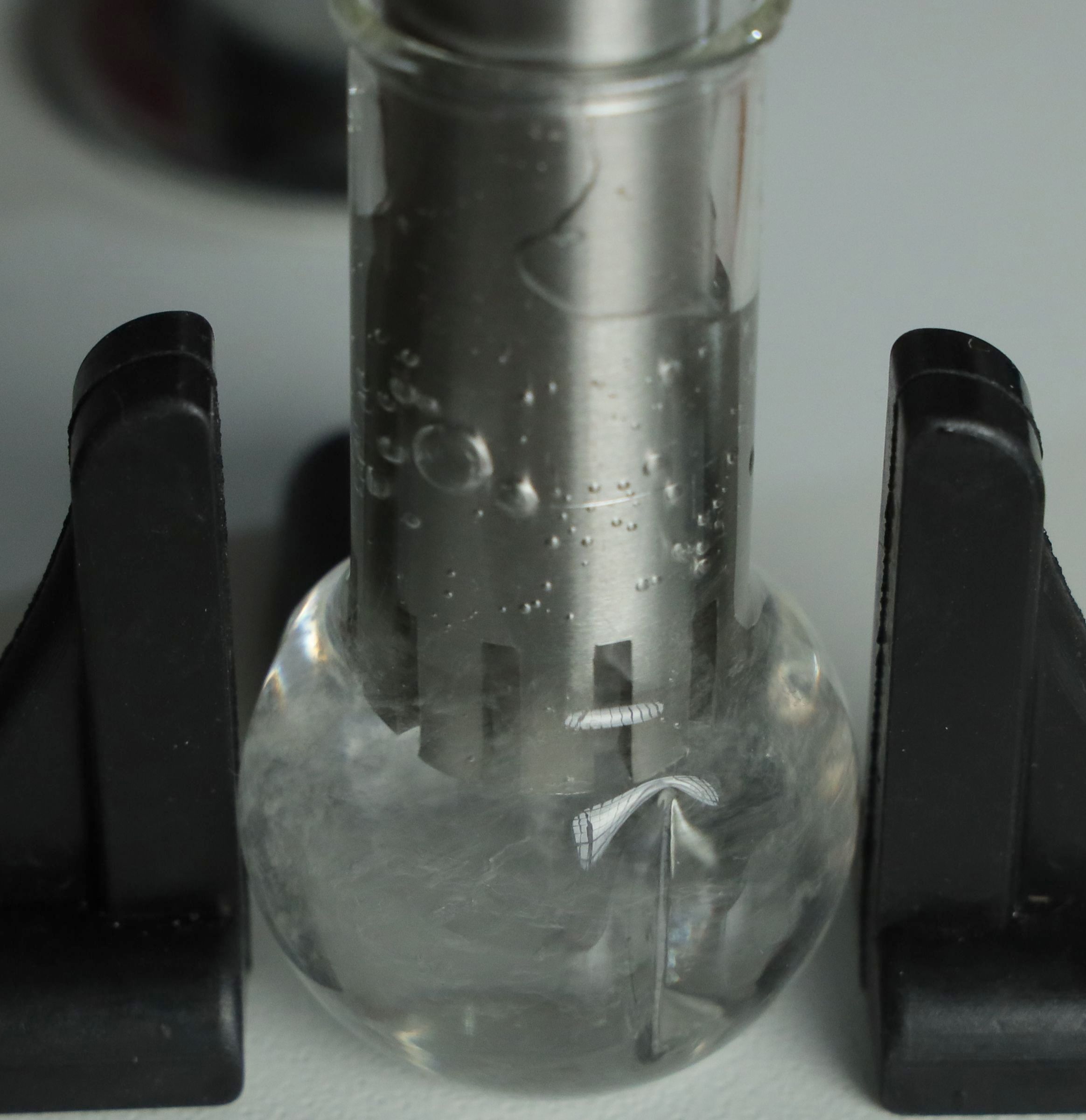}}
\caption{Excitation of 25ml sample by hydrodynamics cavitation with the laboratory homogenizer (FSH-2A) at 10000 rpm. \label{fig:excitation}}
\end{figure}

\textbf{Treatment of the experimental sample.} This work utilizes primarily hydrodynamic cavitation with  a standard laboratory homogenizer (FSH-2A) at 10000 rpm during 5-10 min for 25ml samples, see Fig. \ref{fig:excitation}. The temperature is controlled so that the sample is not heated over 25$^\circ C$. Several attempts have been preformed with ultrasound excitation at 40kHz (in ultrasound cleaner) and at 1.7MHz (with ultrasound vaporizer) also during 5-10 min for 75ml samples; however they essentially increase the temperature of water over 40-45$^\circ C$ and therefore are not used beside preliminary attempts. 

\textbf{Temperature calibration.} Initial temperature of samples was about 18$^\circ C$; it was equalized and set up to 21$^\circ C$. However, during 30 min measurements, the temperature of samples is drifting to the environmental temperature in the laboratory (about 21$^\circ C$-23$^\circ C$) as well as is heating by IR body emission of the experimentator. Thus, a separate series of measurements is conducted to estimate the capillary effects in relation to 25$^\circ C$. The control sample is set to 21$^\circ C$, the experimental sample is set to 25$^\circ C$, the measurement methodology is similar to other measurements, see Table \ref{tab:parameters}. 

\begin{figure}[htp]
\centering
\subfigure[\label{fig:plot1C}]{\includegraphics[width=0.49\textwidth]{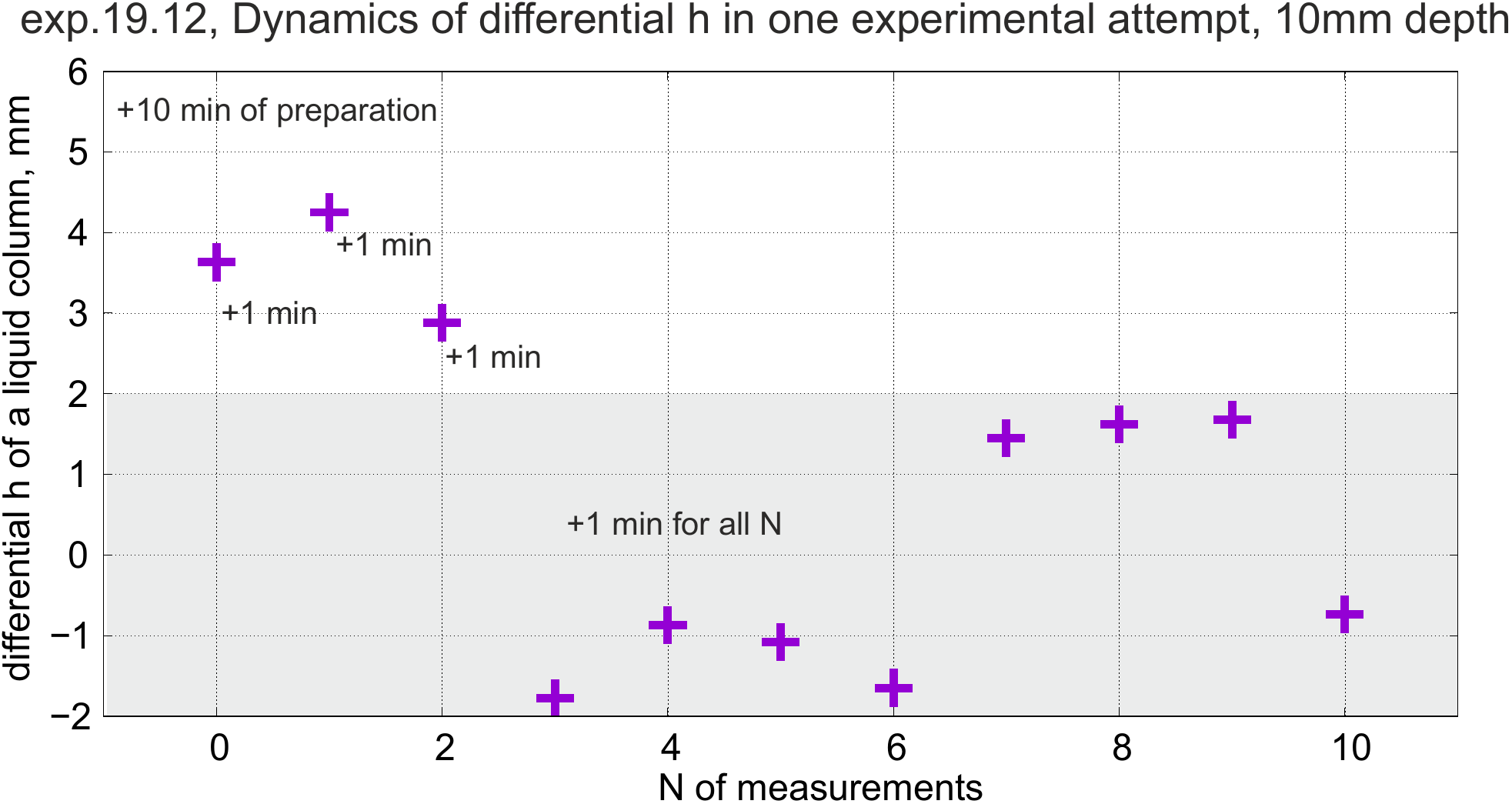}}
\subfigure[\label{fig:plot1A}]{\includegraphics[width=0.49\textwidth]{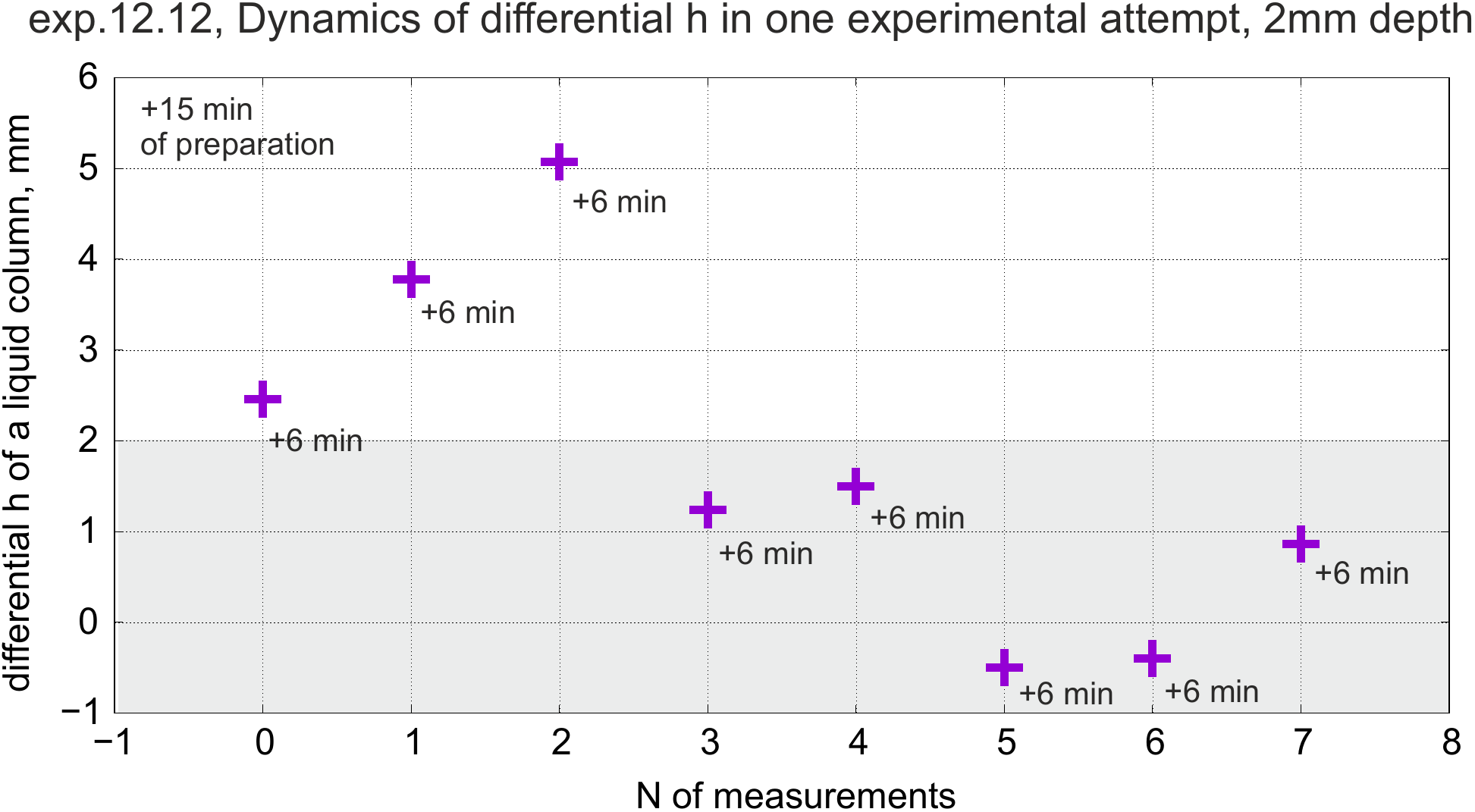}}
\subfigure[\label{fig:plot1B}]{\includegraphics[width=0.49\textwidth]{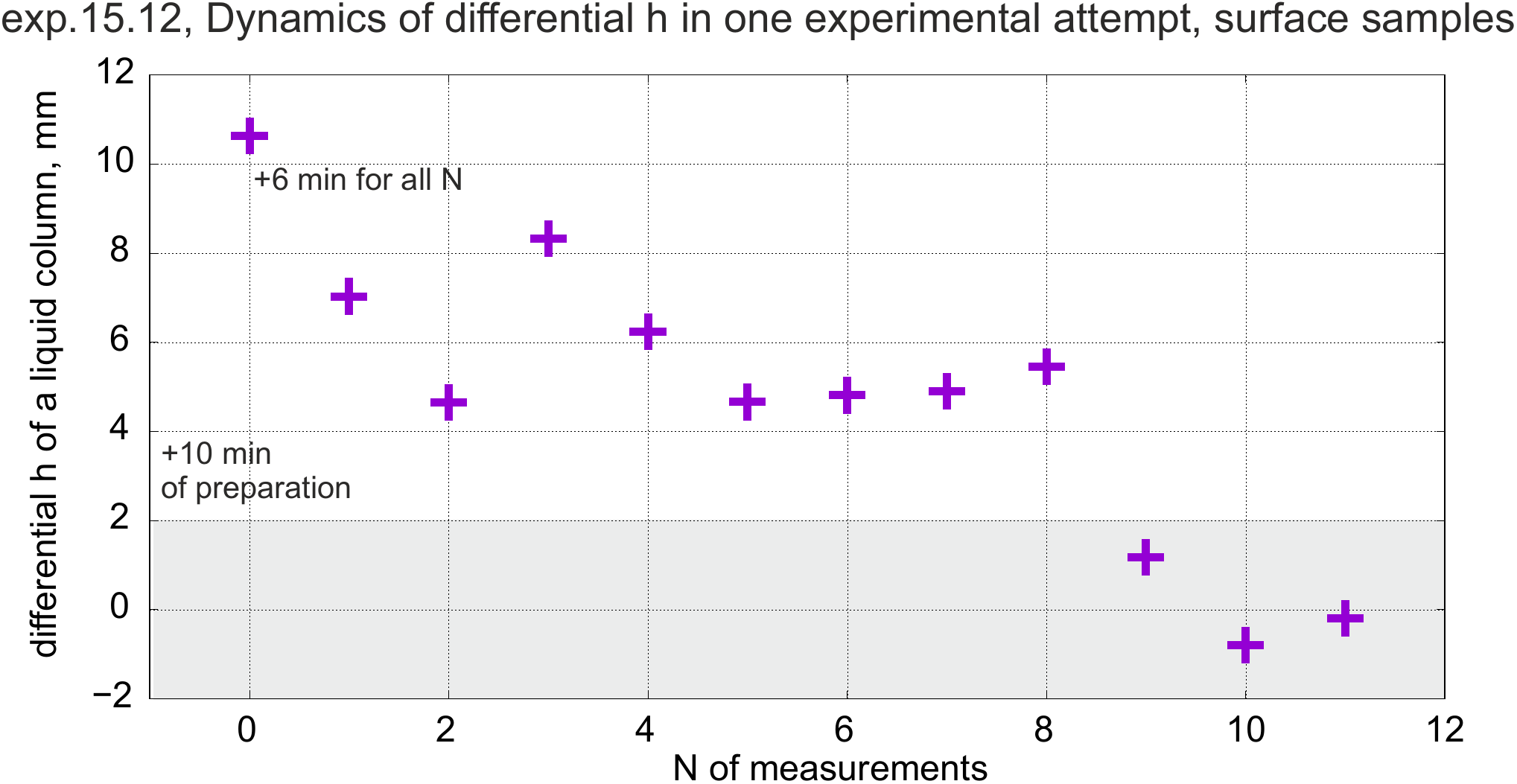}}
\caption{Example of differential $h$ values from one attempt for: \textbf{(a)} 10 mm immersion depth, initial measurements; \textbf{(b)} 2 mm immersion depth, initial measurements; \textbf{(c)} samples taken form surface, 6 min measurements. Essential deviations from control are shown only in a few first measurements of each series. The gray area shows non-significant measurements. \label{fig:plot1}}
\end{figure}

\textbf{Contamination of samples.} The measurements follow good chemical laboratory practice. Two separate sets of all containers made from borosilicate glass were used in preparation of control and experimental samples. However, samples are contacted by excitation devices during treatments. Here, all possible measures were also taken (e.g. periodic cleaning) to avoid contamination of samples. 

\textbf{NMR spectroscopy} of water samples excited by hydrodynamic cavitation was carried out prior to this work with the aid of magnetic resonance image (MRI) scanner Tomikon-S50 (Bruker, 0.5-Tesla, samples 0.5L in plastic containers) and published in \cite{Pershin09NMR} (and shorty in \cite{Pershin22}). This  work compared NMR results with four-photon scattering spectra in the range from -1.5 to 1.5 cm$^{-1}$ for distilled and cavitation water samples and confirmed the enrichment of ortho- isomers by optical measurements. As a reason for the enrichment the work \cite{Pershin09NMR} proposed the supercritical state -- high temperature and pressure of $\sim$10$^{10}$Pa and $\sim$10$^4$K during collapses of cavitation bubbles, resulting in the dissociation of \ce{H_2O} molecules into OH and H radicals and their further recombination with different ortho/para ratio than at the ambient conditions.

\section{Results with 2mm and 10mm immersion depth}

\begin{table*}[h]
\begin{center}
\caption{\small Table of results, the difference $h_{experiment}-h_{control}$ is given as mean$\pm$StDev.\label{tab:parameters}}
\fontsize {9} {10} \selectfont
\begin{tabular}{
p{0.5cm}@{\extracolsep{3mm}}
p{1.5cm}@{\extracolsep{3mm}}
p{2.5cm}@{\extracolsep{3mm}}
p{2.0cm}@{\extracolsep{3mm}}
p{1.0cm}@{\extracolsep{3mm}}
p{1.5cm}@{\extracolsep{3mm}}
p{2.2cm}@{\extracolsep{3mm}}
p{2.0cm}@{\extracolsep{3mm}}
p{2.0cm}@{\extracolsep{3mm}}
}\hline \hline
N & type & N of attempts/ N of measurements & $^\circ C$ control/ experiment & depth, mm & d of tube, mm & initial difference, mm & difference after 3 min, mm & difference after 6 min, mm \\\hline 
1 & control & 4/10           & 21/21                 & 2         & 0.3           & 0.56$\pm$1.11 & 0.29$\pm$0.74 & 0.101$\pm$0.72 \\
2 & control & 2/4            & 21/21                 & 2         & 0.5           & 0.5$\pm$1.22 & -0.16$\pm$0.77 & -0.05$\pm$0.76 \\
3 & control & 2/5            & 21/25                 & 2         & 0.3           & -0.57$\pm$0.53 & -0.35$\pm$0.78 & -0.03$\pm$1.10 \\
4 & cavitation & 3/6         & 21/21                 & 2         & 0.3           & 5.17$\pm$1.18 & 4.47$\pm$0.81 & 4.35$\pm$1.01 \\
5 & cavitation & 3/6         & 21/21                 & surface   & 0.3           & 5.73$\pm$2.81 & 6.63$\pm$1.74 & 7.34$\pm$1.86 \\
6 & cavitation, all samples & 4/19    & 21/21        & surface   & 0.3           & 4.58$\pm$2.09 & 5.26$\pm$1.66 & 5.66$\pm$1.81 \\
\hline \hline 
\end{tabular}
\end{center}
\end{table*}

Preliminary attempts demonstrated sensitivity of measurements to the way capillary tubes are immersed into the fluids. In particular, two effects seemed to be relevant: the immersion depth and stopping the liquid when it enters the capillary tube. To minimize the problem with the initial stoppage of fluids, the capillary tubes are immersed twice, about 1 second apart. Results indicating significant variations due to the fluid retention (i.e. fluids are entering the tubes not at the same time) are discarded from consideration. Each series of measurements is conducted during 30-40 min. Vernier height gauges are calibrated to the initial position of the meniscus in the control fluid, thus each position means the difference between $h$ in control and experimental tubes, see Table \ref{tab:parameters}.

\textbf{Control measurements} have been conducted in different conditions, typically 2-3 measurements per independent excitation attempt. Here, 0.3 and 0.5 tubes demonstrated similar results: 0.101$\pm$0.72 and -0.05$\pm$0.76. Since 0.3mm tubes have better resolution of $h$ dynamics, they are used in further experiments. Changing the temperature of fluids as 21$^\circ C$ (control) to 25$^\circ C$ (experiment) shows at -0.57$\pm$0.53 at the first measurements and -0.03$\pm$1.10 at measurements 6 minutes after. We assume that fluids in capillary tubes equalize the temperature during 6 minutes so that only the initial measurements demonstrate the effect of different temperature. 

\textbf{Experimental measurements with hydrodynamic cavitation} demonstrated essential deviations from the control only in a few first measurements of each series. Figure \ref{fig:plot1A} shows an example of differential $h$ values from one attempt at 2mm immersion depth. Since calculating mean and StDev from all values makes no sense, here we collect only two largest values from each attempt and then represent them in the form of mean$\pm$StDev. Such methodology is well acceptable for qualitative detection strategy with only a few measurements per experiment. 

The results 1-4 from Table \ref{tab:parameters} are obtained with samples taken approximately 2 mm below the surface. Samples taken 10mm below the surface, e.g. by fast immersion, solves the problem of stopping the liquids, however do not demonstrate essential differences in the height of a liquid column. To investigate this issue, Fig. \ref{fig:plot1C} shows a dynamics of differential $h$ values, obtained by fast measurements (about 1min for one measurement) at 10mm depth. We see that the effect is only present for a short period of time (about 3 min) and has a lower amplitude. Taking into account that some samples can be still taken from the near-surface layers of water during a rapid immersion (water is taken into the capillary very quickly), we stopped these attempts.

\begin{figure}
\centering
\subfigure{\includegraphics[width=0.49\textwidth]{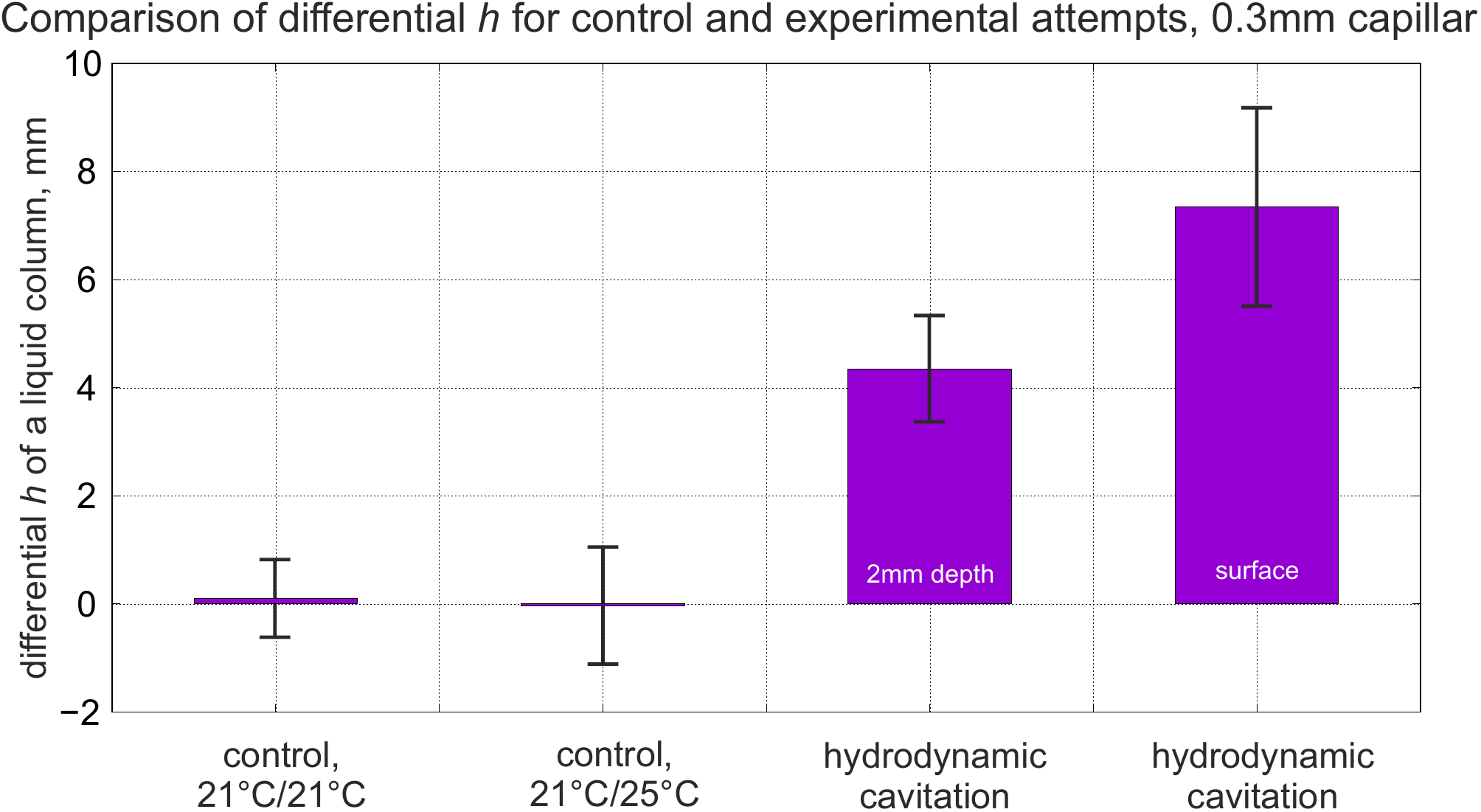}}
\caption{Comparison of control and experimental attempts from Table \ref{tab:parameters} with 0.3mm capillary tubes, six-minute measurements. \label{fig:comparison}}
\end{figure}

Comparison of control and experimental (obtained with the 'two largest' strategy) attempts is shown in Fig. \ref{fig:comparison}.

\section{Results with samples from the surface}

\begin{figure*}
\centering
\subfigure[]{\includegraphics[width=.56\textwidth]{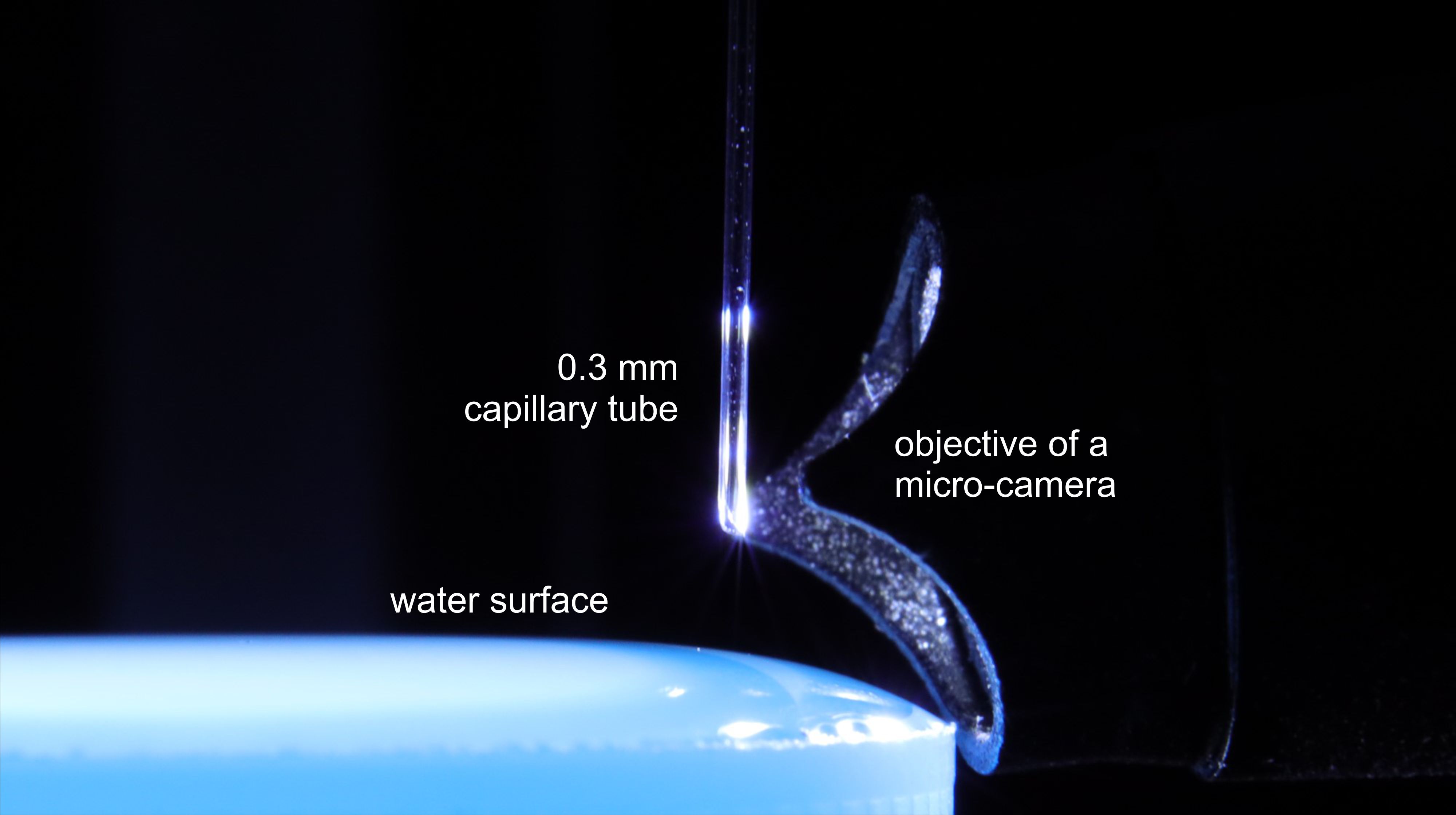}}~
\subfigure[]{\includegraphics[width=.42\textwidth]{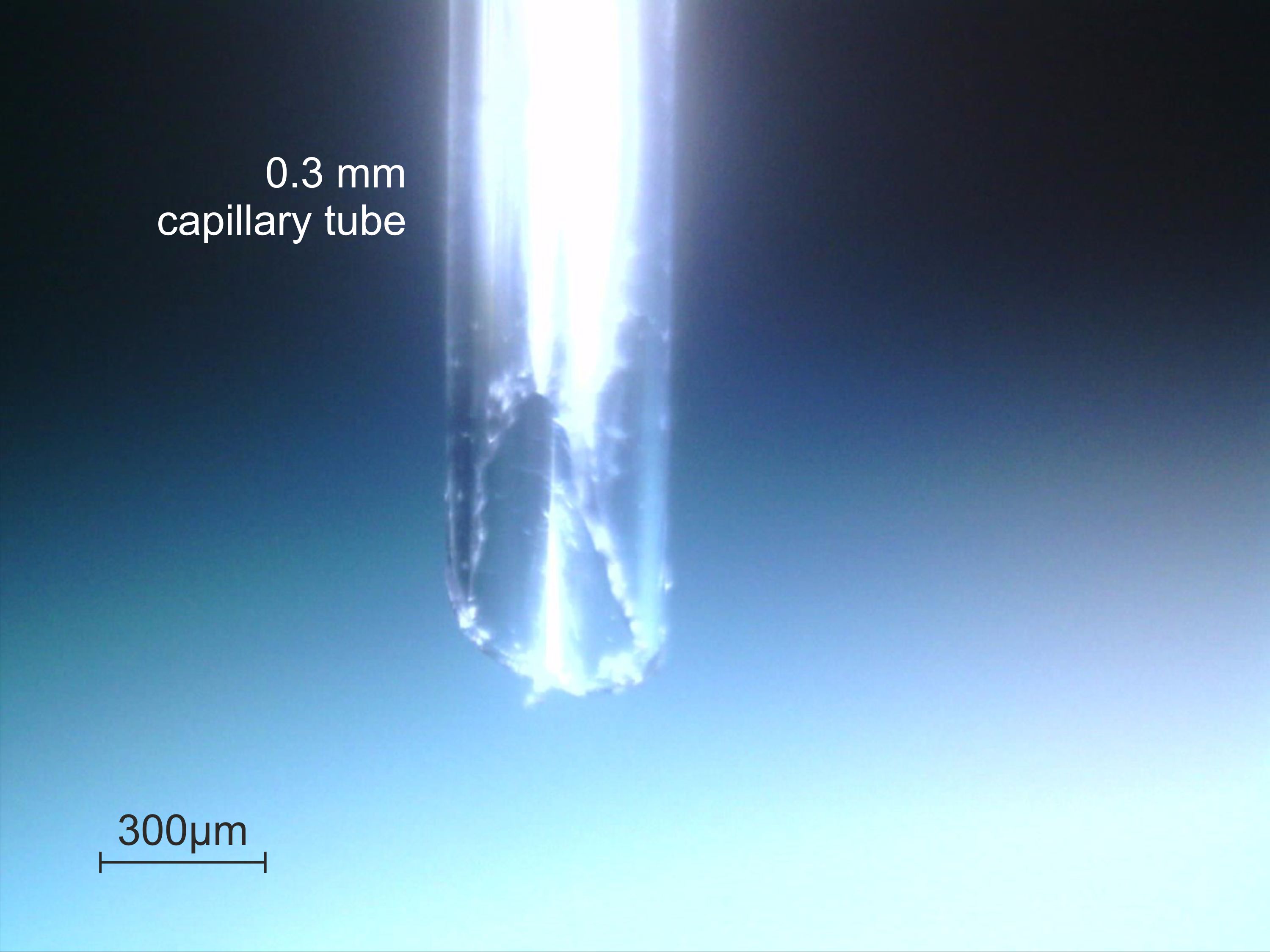}}
\caption{\textbf{(a)} Setup with optical camera for micro-manipulation for taking samples from the surface, shown is the right channel used for control samples; \textbf{(b)} Image from the micro-camera, the same setup as in (a). \label{fig:surface}}
\end{figure*}

Arguments about the existence of ice-like structures at the water interfaces that can potentially lead to a long-term non-equilibrium state of spin isomers are expressed in \cite{Davis12,Pershin12ICE,Odendahl22,Chai09,PershinIce10} within a large discussion in physicochemical community towards ice-like building blocks at air-water, metal-water and hydrophilic interfaces \cite{PhysRevLett.101.036101,Wei01,en8099383,Monserrat20,Novikov10,POULOSE2023814}. Since the thickness of the structured layer is expected to vary between a few \AA~ and a few mm, it is important to have a controllable way for sampling on a micro-scale. To investigate this issue in more detail, we developed the second setup with optical camera for micro-manipulation of tubes, see Fig. \ref{fig:surface}.

\begin{figure}
\centering
\subfigure{\includegraphics[width=0.49\textwidth]{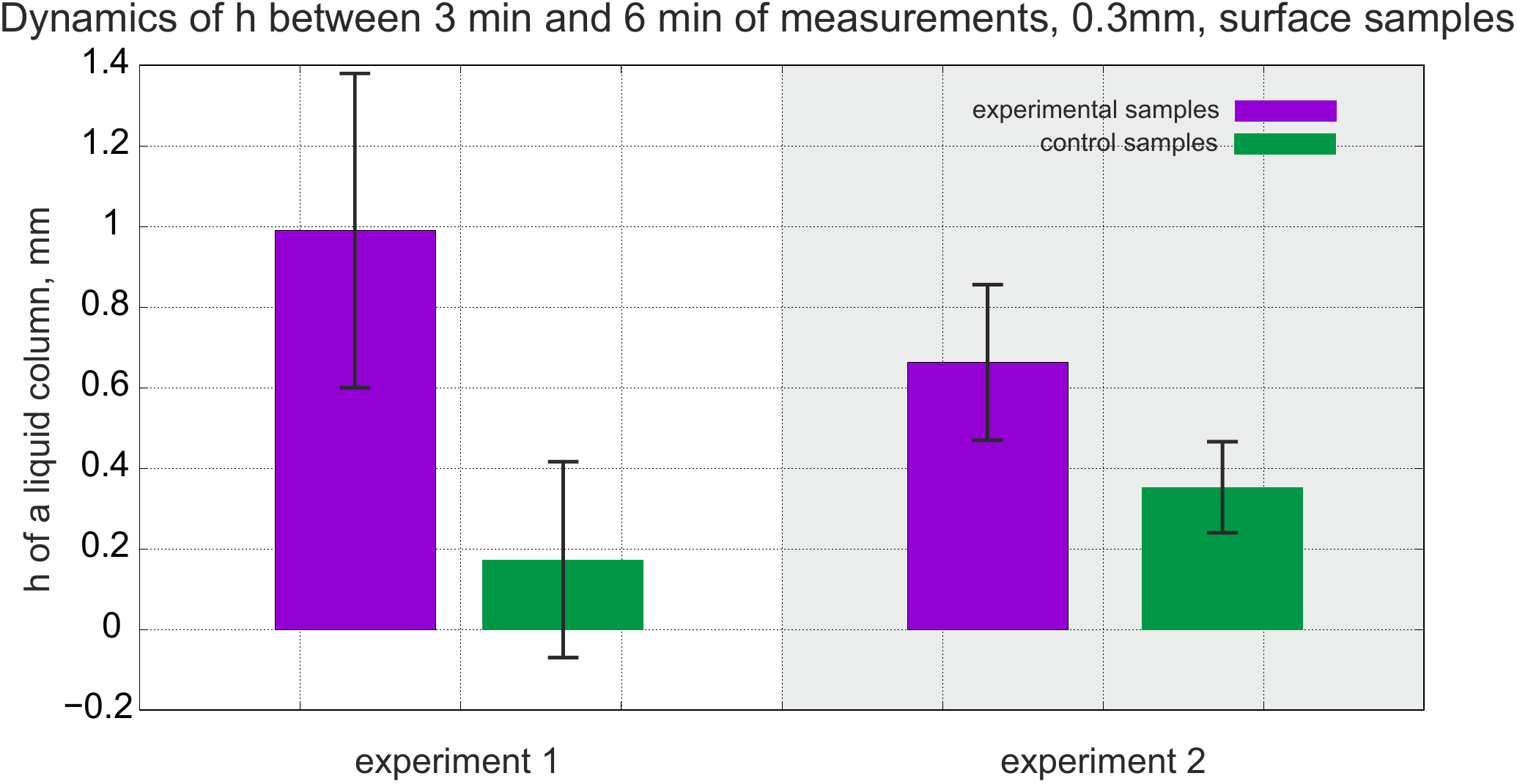}}
\caption{Comparison of $h$ in control and experimental samples between 3min and 6min measurements, the surface-based setup, 0.3mm capillary tubes. \label{fig:speed}}
\end{figure}

\begin{figure*}
\centering
\subfigure[\label{fig:microCameraA}]{\includegraphics[width=0.3\textwidth]{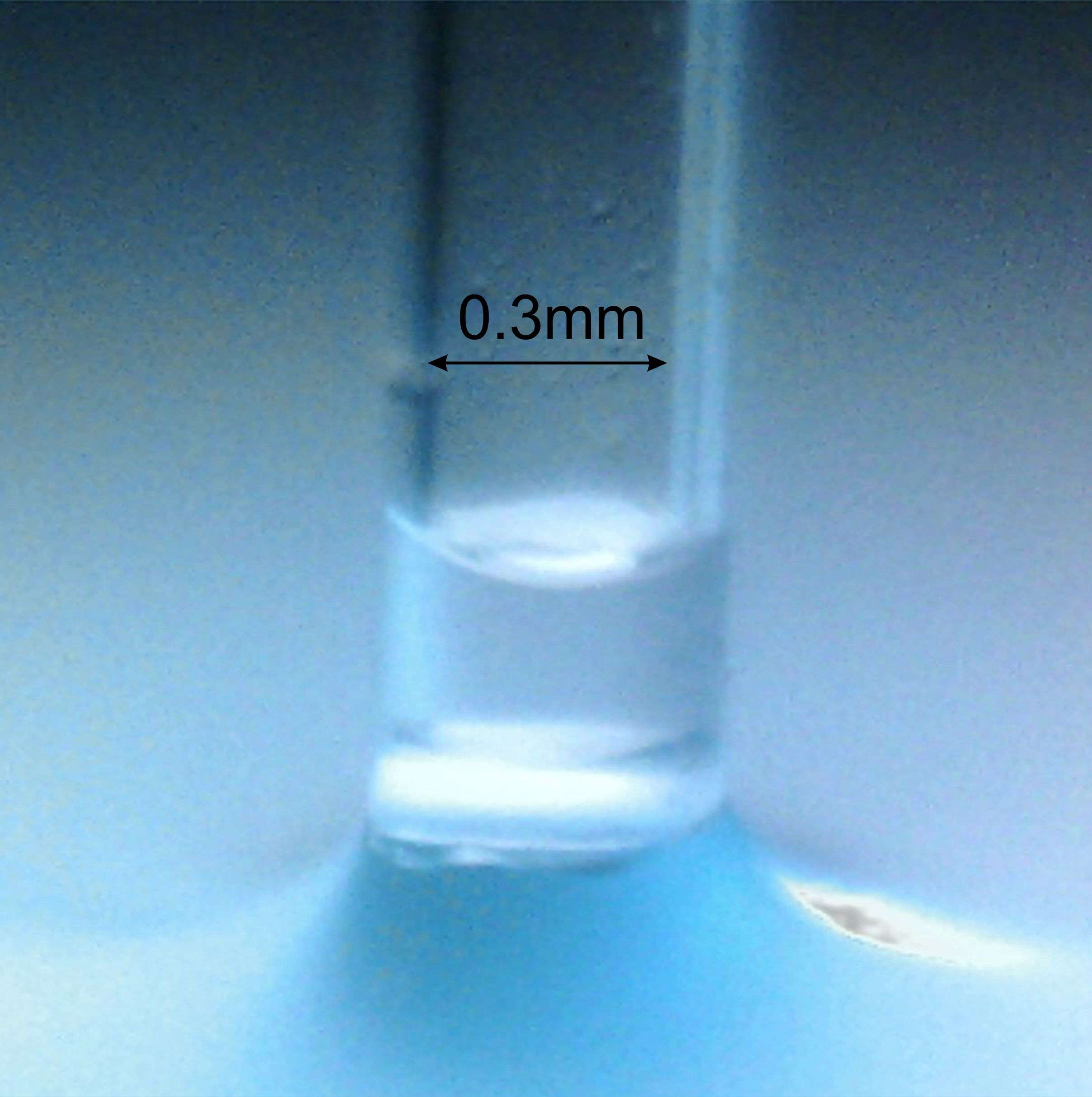}}~
\subfigure[\label{fig:microCameraB}]{\includegraphics[width=0.3\textwidth]{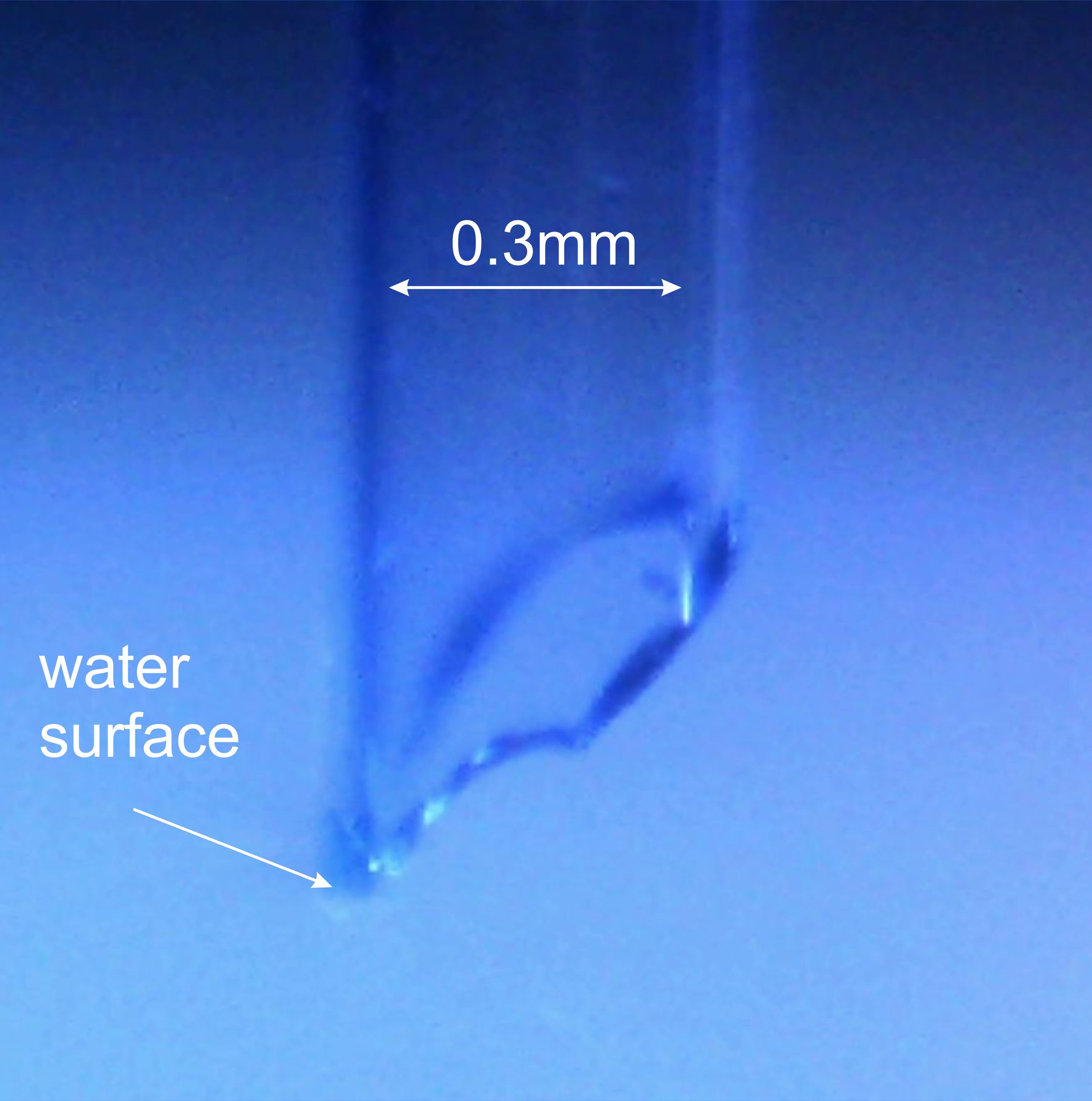}}~
\subfigure[\label{fig:microCameraC}]{\includegraphics[width=0.29\textwidth]{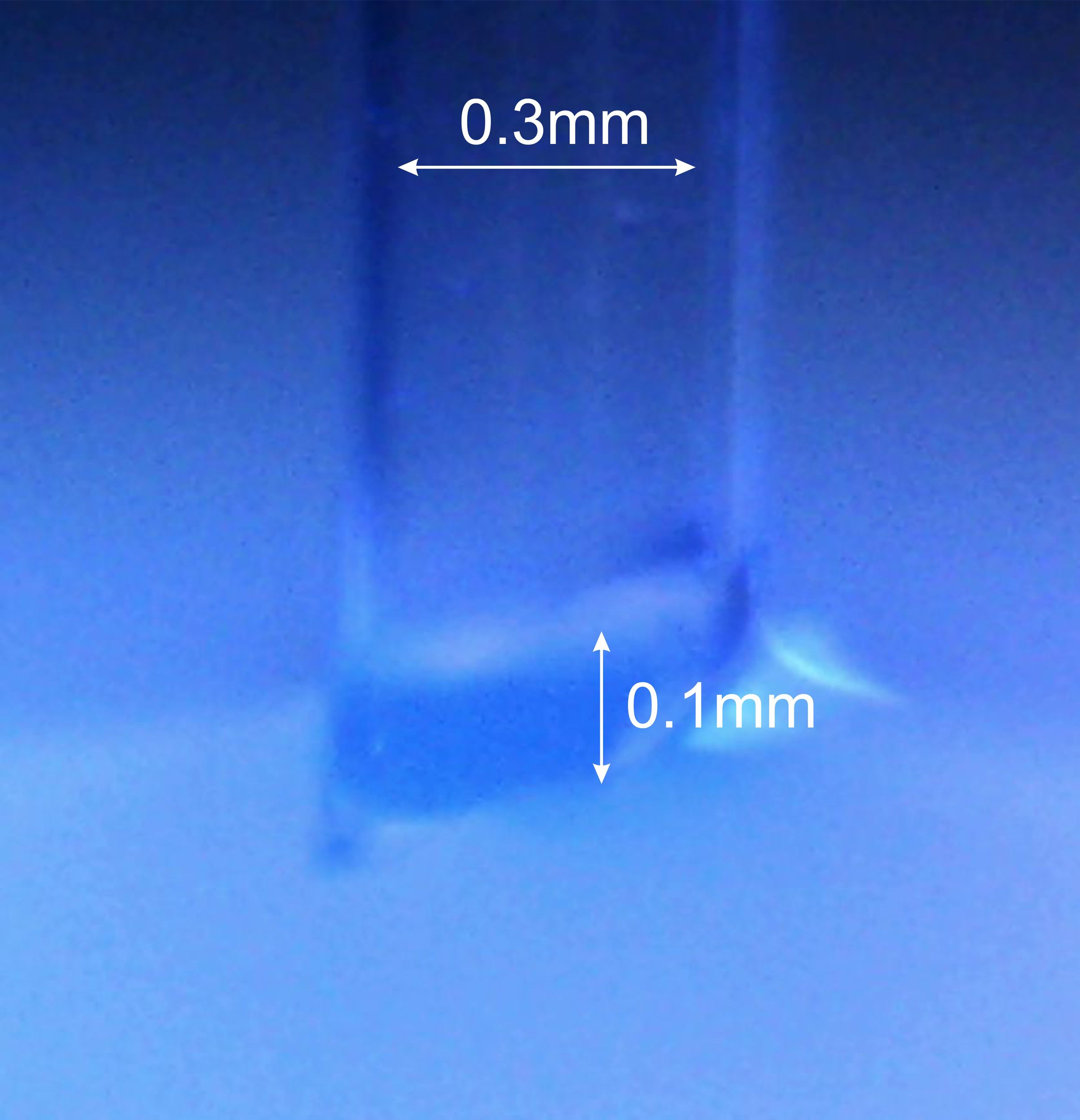}}
\caption{Images from the micro-manipulation camera: \textbf{(a)} Stopping the liquid, typical example of round edges, the meniscus and surrounding surface tension can fluctuate at 0.05-0.1mm level; \textbf{(b)} Typical example of sharp edges; \textbf{(c)} Taking sample from the water surface, estimated depth of water is about 0.1mm. \label{fig:microCamera}}
\end{figure*}

\begin{figure}
\centering
\subfigure{\includegraphics[width=0.49\textwidth]{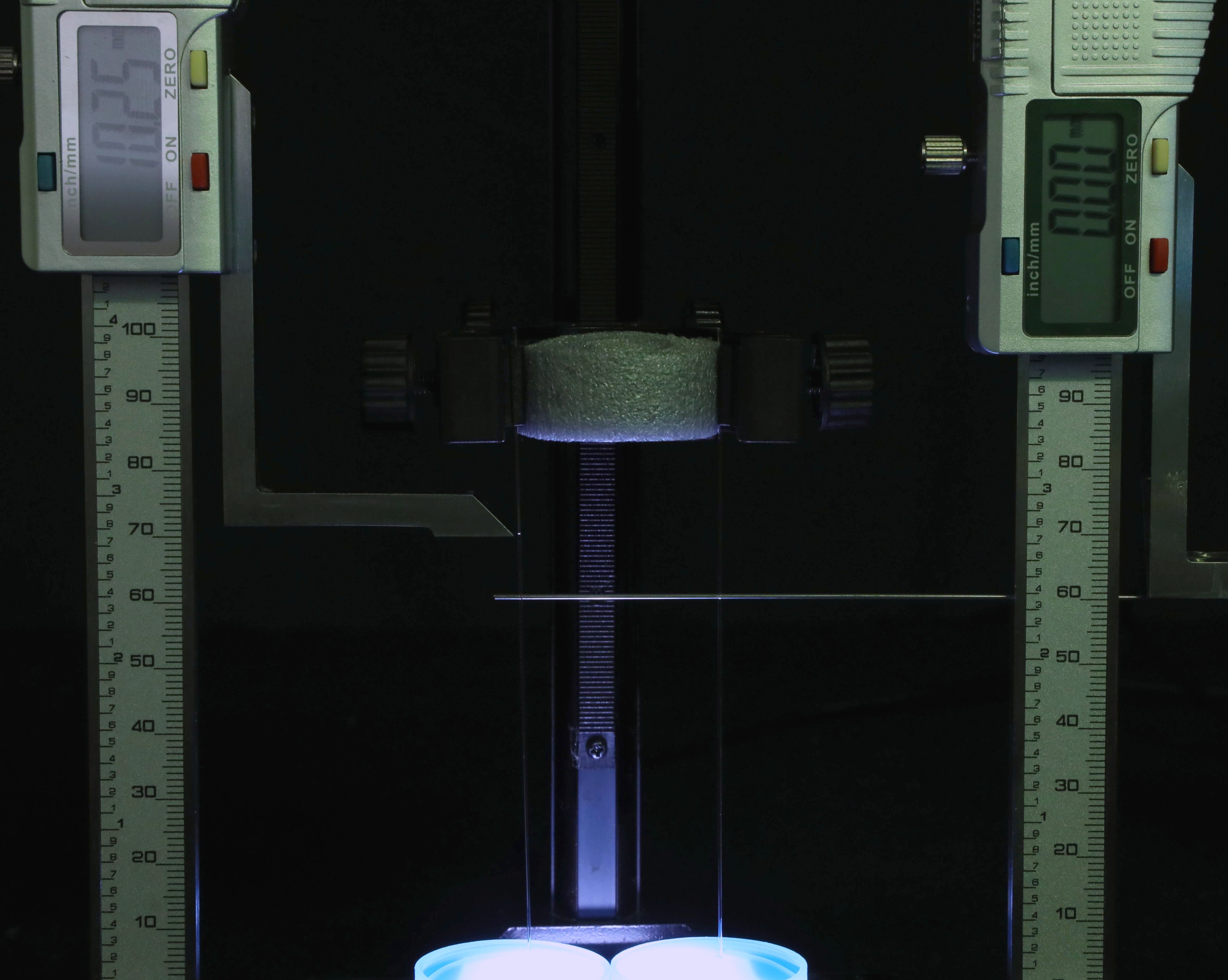}}
\caption{Example of control and experimental measurements with the surface-based setup, the differential $h$ is about 10.25mm, 0.3mm capillary tubes. \label{fig:surface_result}}
\end{figure}

After excitation, the samples were first degassed and thermally balanced in large-area containers and then refilled to flat measuring plates. Vertical alignment of tubes shortly before touching the surface is equalized based on optical feedback so that both tubes contact the surface synchronously. Fig. \ref{fig:microCamera} shows several images from the micro-manipulation camera for the case of sharp-edge and round-edge geometries of tubes, which are significantly different for their further capillary dynamics. In the worst case of sharp-edge geometry (broken edge of a tube), see Fig. \ref{fig:microCameraB}, the samples are taken from about 0.1mm depth, see Fig. \ref{fig:microCameraC}. In the case of round-edge tubes, samples are collected primarily from the surface due to adhesion between fluid and the tube, see Fig. \ref{fig:microCameraA}. Here, the micro-manipulation tool recorded a dynamics of stopping the liquid in round-edge capillary tubes, which was observed mostly in experimental samples. Interesting phenomena consisted in fluctuation (in some moments -- oscillations) of the meniscus and surrounding surface tension at 0.05-0.1mm level. Fluctuations and oscillations of surface tension are also observed in \cite{kernbach2022electrochemical} with electrochemical measurements. We assume that different geometry of tubes contributes to a variation of experimental outcomes. 
 
Results obtained in this setup essentially differ from previous attempts. First, they are stronger in term of differential $h$, here we observed maximal values of over 10mm, see Fig. \ref{fig:surface_result}.
The methodology of collecting results is enhanced, in addition to 'two largest' we used 'all within the effective time' strategy (all values outside the gray region in Figs. \ref{fig:plot1}). This increases the standard deviation, however, this methodology provides better overview over the whole series of measurements. Obtained averaged level is higher then in the case of 2mm depth, e.g. 7.34$\pm$1.86 for 'two largest' and 5.66$\pm$1.81 for 'all within the effective time'. Second, these effects have a longer duration. If the previous attempts typically show no results 30 minutes after excitation, see Fig. \ref{fig:plot1B}, here the results are obtained over a period of about 50-60 minutes, see Fig. \ref{fig:plot1C}. Finally, after 3-minute measurements, control samples significantly delayed their movement, while the experimental samples were still moving. Fig. \ref{fig:speed} compares $h$ between 3min and 6min measurements, we observe about twice the difference.

\section{Discussion}

Considering the average liquid column height of 65 mm in control samples, we can estimate a variation of surface tension in (\ref{eq:liquidColumn}) as 6.7\%-11.3\% up to maximal 15.7\% by hydrodynamic cavitation, assuming that the density of fluid is not changed by this factor. Important is a higher movement velocity of meniscus in experimental samples, shown in Fig. \ref{fig:speed}. These two factors can be used in the qualitative detecting approach.
     
One of the main problems with these experiments concerns the stopping of liquids entering the capillary tube in near-surface layers. Due to variations in the production process (tubes are separated from each other by breaking), sharp edges make it easier for liquid to penetrate, while rounded edges delay it. This creates a variation in the height of a liquid column during six-minute measurements. Considering the fluctuation of the stopped meniscus, the exact mechanism causing this problem is unclear, but it can be attributed to excitation since it occurs in experimental tubes in most cases. We have tested several approaches that should minimize this effect: fast immersion, twice immersing into the fluid or a mechanical destruction of round edges. In replications attempts it would make sense to redesign the measurement scheme to avoid the problem with stopping of liquids.

Samples taken from surface seem to produce more significant results. However here we observed excitation attempts with strong and weak results, although the applied excitation and the preparation approach were identical. In samples obtained from 2mm depth, variations even between measurements were quite essential and limit us to selection of only two largest samples. It can be assumed, spatial concentration of non-equilibrium states has some inhomogeneous character in near-surface layers that involves a random factor into the attempts.

The number of consecutive measurements from an experiment is important. At a depth of 2 mm, only a few initial six-minute measurements show a difference between control and experiment, while for surface samples it is extended to 5-8 measurements. Considering other reports, e.g. \cite{Otsuka06}, which deals with the dynamics of surface tension after different excitations, we also see about 30 minutes strong and 30 minutes weak relaxation time. Allowing for about 10-15 minutes for degassing and temperature equilibration, our measured relaxation time agrees well with these results. In other words, the observed surface tension changes caused by excitation have a time limit of 30–60 minutes.

\section{Conclusion}

Attempts in this work have been conducted with excitation of fluidic samples by hydrodynamic cavitation, which has been reported to introduce a non-equilibrium state of spin isomers and enrich the samples by 12\%-15\% of ortho- isomer. Experiments and measurements based on capillary effects in 0.3mm tubes demonstrated a variation of surface tension as 6.7\%-11.3\% up to maximal 15.7\%, which can be also tracked to a spin conversion of water isomers in ice-like structures at the water interface. The near-surface effects can be confirmed to a certain extent, as we observed weakening the results with immersion depth, taking into account that capillary tubes penetrate different water layers during a single immersion. Considering NMR \cite{Pershin09NMR}, electrochemical \cite{kernbach2022electrochemical}, thermal \cite{kernbach23Thermal} and optical \cite{Pershin14,Pershin09Temp} measurements, we have arguments to consider different water interfaces \cite{Chai09,Pershin12ICE,Odendahl22} as candidates for the long-term preservation of non-equilibrium states of spin isomers in the liquid phase. The observed effect will allow studying in more detail the transport of water enriched with ortho-isomers of \ce{H_2O} through nanocapillaries of membrane aquaporin channels with a diameter of 3\AA \cite{Murata00,Agre06}.

A serious obstacle that hinders the further development of more accurate measuring devices is the variation of results. On the one hand, there are manufacturing reasons that vary technological parameters of capillary tubes. On the other hand, we see an inhomogeneous spatial concentration of non-equilibrium states as well as some random factors during the excitation attempts.

Obtained results suggest that this approach can be a fast and inexpensive qualitative detector of spin-based phenomena affecting the surface tension. As an extension of these experiments we conducted measurements with the root system and sap flow movement of hydroponic plants irrigated with excited water. These mechanisms also have a capillary character, we observed a number of anomalies in biological parameters of such plants. With a view to future research, the increasing accuracy towards measuring devices and biological applications represent two interesting aspects that should be pursued further.

\section{Acknowledgements}

The first author of this work is supported by EU-H2020 Project 'WATCHPLANT: Smart Biohybrid Phyto-Organisms for Environmental In Situ Monitoring', grant No: 101017899 funded by European Commission.

\small
\IEEEtriggeratref{20}
\bibliographystyle{elsarticle-num}


\end{document}